\documentclass{svjour3}                     
\smartqed  
\usepackage{graphicx}
\usepackage{amssymb,amsmath}
\begin{document}

\title{The conservation of energy-momentum and the mass for the graviton
}


\author{M.E.S.Alves \and O.D.Miranda \and J.C.N.de Araujo}


\institute{M.E.S.Alves \and O.D.Miranda \and J.C.N.de Araujo \at
              Divis\~ao de Astrof\'isica, Instituto Nacional de Pesquisas Espaciais,\\
              Avenida dos Astronautas 1758, S\~ao Jos\'e dos Campos 12227-010 SP, Brazil\\
              \email{alvesmes@das.inpe.br}           
           \and
           O.D.Miranda \at
              oswaldo@das.inpe.br
       \and
       J.C.N. de Araujo \at
          jcarlos@das.inpe.br
}

\date{Received: date / Accepted: date}

\maketitle

\begin{abstract}
In this work we give special attention to the bimetric theory of
gravitation with massive gravitons proposed by Visser in 1998. In
his theory, a prior background metric is necessary to take in
account the massive term. Although in the great part of the
astrophysical studies the Minkowski metric is the best choice to
the background metric, it is not possible to consider this metric
in cosmology. In order to keep the Minkowski metric as background
in this case, we suggest an interpretation of the energy-momentum
conservation in Visser's theory, which is in accordance with the
equivalence principle and recovers naturally the special
relativity in the absence of gravitational sources. Although we do
not present a general proof of our hypothesis we show its validity
in the simple case of a plane and dust-dominated universe, in
which the `massive term' appears like an extra contribution for
the energy density.

\end{abstract}

\section{Introduction}
\label{intro}
Could the graviton have a non-zero rest mass? The observations have shown that this is a possibility. One of the most accurate bounding on the mass of the graviton comes from the observations of the planetary motion in the solar system. Variations on the third Kepler law comparing the orbits of Earth and Mars can lead us to $m_g < 7.8 \times 10^{-55}g$ \cite{Talmadge88}. Another bound comes from the analysis of galaxy clusters that lead to $m_g <  2 \times 10^{-62}g$ \cite{Goldhaber74} which is considerably more restrictive but less robust due to the uncertainties in the content of the universe in large scales. Studying rotation curves of galactic disks, \cite{JC} has found that we should have a massive graviton of $m_g \ll 10^{-59}g$ in order to obtain a galactic disk with a scale length of $b\backsim10$ kpc.

The above tests are obtained from static fields based on deviations of the newtonian gravity. In the weak field limit has been proposed \cite{Finn2002} to constraint $m_g$ using data on the orbital decay of binary pulsars. From the  binary pulsar  PSR B1913+16 (Hulse-Taylor pulsar) and PSR B1534+12 it is found the limit $m_g < 1.4 \times 10^{-52}g$, which is weaker than the bounds in static field.

It is worth recalling that the mass term introduced via a Pauli-Fierz (PF) term in the linearized approximation produces a theory whose predictions do not reduce to those of general relativity for $m_g \rightarrow 0$. This is the so called van Dam Veltmann Zakharov discontinuity \cite{Veltman1970}. Moreover the Minkowski space as background metric is unstable for the PF theory \cite{Gruzinov2005}. However, there is no reason to prefer the PF term over any other non-PF quadratic terms.

It is important to emphasize that these mass terms do not have clear extrapolation to strong fields. A way to do that was proposed by Visser \cite{visser98}. To generalize the theory to strong fields, Visser makes use of two metrics, the dynamical metric ($g_{\mu\nu}$) and a non-dynamical background metric ($\left( g_0\right) _{\mu\nu}$) that are connected by the mass term. Although adding a prior geometry is not in accordance with the usual foundations underlying Einstein gravity, it keeps intact the principles of equivalence (at least in its weak form) and general covariance in the Visser's work. Some interesting physical features emerge from the theory such as extra states of polarizations of the gravitational waves \cite{wayne2004}.

In the present article, we explore some aspects which are not treated by Visser in his original paper. In the great part of the astrophysical studies the Minkowski metric is the most appropriate choice to the background metric. However, in the study of cosmology, it is not possible to consider this kind of metric, and we need some prior considerations regarding a background metric. Once this problem emerges from the coupling of the two metrics and the energy's conservation condition, we analyze an alternative interpretation of this condition. We also show that this interpretation is in accordance with the equivalence principle and recovers naturally the special relativity in the absence of gravitational sources. Arguments in favor of a Minkowskian background metric in Visser's theory are also considered.

This paper is organized as follows: in section \ref{sec:1} we show how to introduce a mass for the graviton through a non-PF term. We present the strong field extrapolation as given by Visser in section \ref{sec:2}. In section \ref{sec:3} we show that the theory is not in accordance with a Minkowski background metric in the study of cosmology. In section \ref{sec:4} we re-interpret the stress-energy conservation in order to keep Minkowski as background in any case. In particular, we show that our re-interpretation is in accordance with the equivalence principle. In section \ref{sec:5} we show why Minkowski is the most natural choice to the background metric. We briefly study some cosmological consequences of our interpretation of the energy-momentum conservation in section \ref{sec:6}. And finally, we present our conclusions in the last section.

\section{The linearized approximation}
\label{sec:1}
The action of a massive gravity in weak field limit may be given by
\begin{equation}\label{actionweak}
I=\int d^4x \bigg\{ \frac{1}{2}\left[h^{\mu\nu}{\Box}^2 h_{\mu\nu}
-\frac{1}{2}h\Box^2h\right] - \frac{1}{2}\frac{m_g^2c^2}{\hbar^2}\left[ h^{\mu\nu}h_{\mu\nu}-\frac{1}{2}h^2\right]+\frac{8\pi G}{c^4}h^{\mu\nu}T_{\mu\nu} \bigg\} ,
\end{equation}
where the first term is the linearization of the usual Einstein-Hilbert Lagrangian and the second term is the mass term for the graviton that is a non-PF one. This fact is essential to have a well-behaved classical limit as the graviton mass goes to zero. From equation (\ref{actionweak}) we have the field equation in the weak field regime
\begin{equation}\label{weakfieldeq}
\Box^2\left[ h_{\mu\nu}-\frac{1}{2}\eta_{\mu\nu}h\right]-\frac{m_g^2c^2}{\hbar^2}\left[ h_{\mu\nu}-\frac{1}{2}\eta_{\mu\nu}h\right]=-\frac{16\pi G}{c^4}T_{\mu\nu}
\end{equation}
or
\begin{equation}\label{weakfieldtwo}
\left( \Box^2-\frac{m_g^2c^2}{\hbar^2}\right) \bar{h}_{\mu\nu}=-\frac{16\pi G}{c^4}T_{\mu\nu}
\end{equation}
where
\begin{equation}
\bar{h}_{\mu\nu}=h_{\mu\nu}-\frac{1}{2}\eta_{\mu\nu}h
\end{equation}

The equation (\ref{weakfieldtwo}) is a Klein-Gordon type. Note that this equation in the limit $m_g \rightarrow 0$ gives us the weak field equations as in general relativity and the newtonian potential in the non-relativistic limit.

Taking the mass term as above we obtain the condition
\begin{equation}\label{gauge}
\partial_\nu\bar{h}^{\mu\nu}=0.
\end{equation}
as a natural consequence of the energy's conservation \cite{visser98}, instead of a gauge condition as in general relativity. But, as we will see later, this is not the case when one considers strong fields.

\section{The Visser's strong field equations}
\label{sec:2}
Following Visser, the extrapolation of the mass term in the equation (\ref{actionweak}) to strong fields could be made by introducing a background metric $g_0$, which would not be subject to a dynamical equation. So, the mass term of strong fields is given by the action
\begin{eqnarray}\label{actiostrong}
I_{mass} &=& \frac{1}{2}\frac{m_g^2c^2}{\hbar^2}\int d^4x\sqrt{-g_0}\bigg\{( g_0^{-1})^{\mu\nu}( g-g_0)_{\mu\sigma}( g_0^{-1})^{\sigma\rho}( g-g_0)_{\rho\nu} \nonumber
\\ & & -\frac{1}{2}\left[( g_0^{-1})^{\mu\nu}( g-g_0)_{\mu\nu}\right]^2\bigg\}
\end{eqnarray}
that recovers the action (\ref{actionweak}) when we consider the weak field limit:
\begin{equation}
g_{\mu\nu}=(g_0)_{\mu\nu}+h_{\mu\nu},~~~|h|<<1.
\end{equation}

Then, the full action considered by Visser is
\begin{equation}\label{fullaction}
I=\int d^4x\bigg[ \sqrt{-g}\frac{c^4R(g)}{16\pi G}+{\cal{L}}_{mass}(g,g_0)+{\cal{L}}_{matter}(g)\bigg]
\end{equation}
in which the background metric shows up only in the mass term for the graviton. The equations of motion that comes from (\ref{fullaction}) may be written such as the Einstein equations
\begin{equation}\label{eqmotion}
G_{\mu\nu}=-\frac{8\pi G}{c^4}\left[ T_{\mu\nu}+T_{\mu\nu}^{mass}\right],
\end{equation}
where the contribution of the mass term appears like an extra contribution to the stress-energy tensor, namely
\begin{eqnarray}\label{extrastress-energy}
T^{\mu\nu}_{mass} &=& -\frac{m_g^2c^6}{8\pi G\hbar^2} \bigg\{ \left( g_0^{-1}\right)^{\mu\sigma}\big[ \left( g-g_0\right)_{\sigma\rho} \nonumber
\\ & & -\frac{1}{2}(g_0)_{\sigma\rho}\left( g_0^{-1}\right) ^{\alpha\beta}\left( g-g_0\right) _{\alpha\beta}\big]\left( g_0^{-1}\right)^{\rho\nu}\bigg\}.
\end{eqnarray}

Following equation (\ref{gauge}), the natural extrapolation to strong fields is
\begin{equation}\label{massconserv}
\nabla_\nu T^{\mu\nu}_{mass} = 0.
\end{equation}

\section{Visser's field equations with a Minkowski background metric}
\label{sec:3}
As pointed out by Visser \cite{visser98}, the most sensible choice for almost all astrophysical applications is to choose $g_0$ as Minkowski. However, some problems appear when we consider this kind of background in cosmology.

To show how these problems emerges, we take the Robertson-Walker as dynamical metric and we consider $k=0$ for simplicity:
\begin{equation}\label{frw}
ds^2=c^2dt^2-a^2(t)\left[ dr^2+r^2(d\theta^2+sin^2\theta d\phi^2)\right].
\end{equation}

To the background metric, we take the following class of metrics:
\begin{equation}\label{backmetric}
ds_0^2=b^2_0(t)c^2dt^2-a^2_0(t)\left[ dr^2+r^2(d\theta^2+sin^2\theta d\phi^2)\right].
\end{equation}

Using these two metrics in the mass tensor and applying (\ref{massconserv}) we obtain
\begin{eqnarray}\label{aoboa}
\frac{\dot{a}}{a}\bigg[\bigg(\frac{a}{a_0b_0}\bigg)^2+\frac{1}{4}\bigg(\frac{a}{a_0}\bigg)^4+\frac{1}{4b_0^4}-\frac{1}{b_0^2}\bigg]+\frac{1}{2}\frac{\dot{a}_0}{a_0}\bigg(\frac{a}{a_0b_0}\bigg)^2 \nonumber \\ -\frac{\dot{b}_0}{b_0}\bigg[\frac{1}{2}\bigg(\frac{a}{a_0b_0}\bigg)^2+\frac{1}{3b_0^4}-\frac{2}{3b^2_0}\bigg]=0,
\end{eqnarray}
where dots represent time derivatives.

Thus, $a_0(t)$, $b_0(t)$ and the scale factor $a(t)$ are related to by the differential equation (\ref{aoboa}). For example, if we choose the background metric as Minkowski ($a_0=b_0=1$), we obtain that the dynamical metric is Minkowski too. Obviously this is not the case in an expanding Universe, for example.

So, in this case, we cannot consider Minkowski and we need some particular choice to the background metric. Some of these possible choices are discussed in the Visser's paper \cite{visser98}.

If a consistent gravitation theory is based on a prior metric, we expect that such a metric would be compatible with any astrophysical case. Once the problems regarding the Minkowskian background metric arises from the condition (\ref{massconserv}), we will explore an alternative interpretation of the energy-momentum conservation in the remaining of the paper. Such a interpretation has the intention of to keep Minkowski as the background metric in any astrophysical study in Visser's theory.

\section{The energy-momentum conservation revisited}
\label{sec:4}
From the field equations in the Visser's theory we may adopt an alternative energy-momentum conservation condition. Taking the divergence of (\ref{eqmotion}), the left-hand-side is a Bianchi identity that is automatically null and from the right-hand-side we get
\begin{equation}\label{new conservation}
\nabla_\nu\left[ T^{\mu\nu}+T_{mass}^{\mu\nu}\right]=0
\end{equation}
We will verify if this equation is in accordance with the equations of motion of a free fall test particle describing a geodesic and, therefore, if it is in accordance with the equivalence principle. In the well known Rosen bimetric theory of gravitation \cite{rosen1973}, for example, it was pointed out the importance of the field equations be in accordance with the geodesic equation which is obtained independently.

To proceed, we adopt the energy momentum tensor to a perfect-fluid:
\begin{equation}\label{perfect fluid}
T^{\mu\nu}=(\rho+p)U^\mu U^\nu+pg^{\mu\nu}.
\end{equation}
Substituting this into equation (\ref{new conservation}) we have
\begin{equation}
\left[(\rho+p)U^\mu U^\nu+pg^{\mu\nu}\right]_{;\nu} =-T^{\mu\nu}_{mass~;\nu}
\end{equation}
\begin{equation}\label{cons mass}
\left[ (\rho+p)U^\nu\right]_{;\nu}U^\mu+(\rho+p){U^\mu}_{;\nu}U^\nu=-T^{\mu\nu}_{mass~;\nu}
\end{equation}
where ``$;$ " denotes the covariant derivative.

Multiplying (\ref{cons mass}) by $U_\mu$ and using
\begin{equation}\label{proper vel}
U^\mu U_\mu=1,
\end{equation}
we obtain
\begin{equation}\label{cons mass 2}
\left[ (\rho+p)U^\nu\right]_{;\nu}+(\rho+p)U_\mu {U^\mu}_{;\nu}U^\nu=-T^{\mu\nu}_{mass~;\nu}U_{\mu}.
\end{equation}

Manipulating ($\ref{proper vel}$) we have
\begin{equation}\label{proper vel 2}
{U^\mu}_{;\nu}U^{\nu}=-U^\mu {U^\nu}_{;\nu};
\end{equation}
from which we can rewrite equation (\ref{cons mass 2}) as
\begin{equation}\label{cons mass 3}
\left[ (\rho+p)U^\nu\right]_{;\nu}-(\rho+p)U^\mu U_\mu {U^\nu}_{;\nu}=-T^{\mu\nu}_{mass~;\nu}U_{\mu}.
\end{equation}

From (\ref{proper vel}) we can find that the second term in the left-hand-side of (\ref{cons mass 3}) is zero, therefore
\begin{equation}\label{cons mass 4}
\left[ (\rho+p)U^\nu\right]_{;\nu}=-T^{\mu\nu}_{mass~;\nu}U_{\mu}.
\end{equation}

Now substituting (\ref{cons mass 4}) in (\ref{cons mass}) we get
 \begin{equation}
-U_{\alpha}T^{\alpha\nu}_{mass~;\nu}U^{\mu}+(\rho+p) {U^\mu}_{;\nu}U^\nu=-T^{\mu\nu}_{mass~;\nu}.
\end{equation}

Since the strong field equations are in accordance with the geodesic equation we have
\begin{equation}
{U^\mu}_{;\nu}U^\nu=0
\end{equation}
which can be rewritten as
\begin{equation}
\frac{d^2x^\mu}{d\tau^2}+\Gamma^\mu_{\alpha\nu}\frac{dx^\alpha}{d\tau}\frac{dx^\nu}{d\tau}=0.
\end{equation}

The last equation can be obtained independently by considering a free fall test particle and the equivalence principle, just as the general relativity theory. Therefore, we conclude that since the energy-momentum conservation condition (\ref{new conservation}) is in accordance with the equivalence principle, the following relation to the mass term needs to be respected:
\begin{equation}\label{mass term condition}
T^{\mu\nu}_{mass~;\nu}=U^{\mu}U_\alpha T^{\alpha \nu}_{mass~;\nu}.
\end{equation}

If we adopt the four-velocity in the rest frame
\begin{equation}\label{propervelnew}
U_\mu=(1,0,0,0),
\end{equation}
then, we will need to have non null components of the divergence of the mass term when $\mu=0$ and $\nu=0,1,2,3$.

Note that the condition imposed for the mass term (\ref{mass term condition}) is not dependent on the form of the tensor $T^{\mu\nu}_{mass}$, so the expression (\ref{new conservation}) is valid to any second rank tensor ``interacting" with the perfect fluid.

\section{Arguments in favor of a Minkowski background}
\label{sec:5}
A classical theory of gravity with a massive graviton apparently needs a background metric for the propagation of this particle. But what is the best physical choice to a background metric? In the Rosen theory \cite{rosen1973} the second metric is a flat metric that describes the inertial forces. We will analyze this issue in Visser's theory.

To do that, we take the field equations (\ref{eqmotion}) in the absence of gravitational source:
\begin{equation}
G^{\mu\nu}=\frac{8\pi G}{c^4}T^{\mu\nu}_{mass}.
\end{equation}
In this particular case, following the treatment that we give in this paper, the covariant divergence produces:
\begin{equation}\label{vac cons}
\nabla_\nu T^{\mu\nu}_{mass}=0.
\end{equation}
Once the mass tensor is constructed by the dynamical metric and by the background metric (and not by derivatives of the metrics), we can conclude that the \textit{most simple way} of satisfying (\ref{vac cons}) is:
\begin{equation}\label{nulity of g0}
\nabla_\nu(g_0)^{\mu\nu}=0
\end{equation}
since the divergence of $g_{\mu\nu}$ is null by construction of the covariant derivatives. Then, the natural solution of (\ref{nulity of g0}) is:
\begin{equation}\label{metrics equality}
(g_0)_{\mu\nu}=g_{\mu\nu}.
\end{equation}
Which by the construction of the mass term (\ref{extrastress-energy}) leads to
\begin{equation}
T^{\mu\nu}_{mass}=0
\end{equation}
and therefore
\begin{equation}\label{vac sol}
G_{\mu\nu}=0.
\end{equation}

In the absence of gravitational sources the simplest solution of (\ref{vac sol}) is:
\begin{equation}
g_{\mu\nu}=\eta_{\mu\nu}
\end{equation}
where $\eta_{\mu\nu}$ is the Minkowski metric and by (\ref{metrics equality}) we get:
\begin{equation}
(g_0)_{\mu\nu}=\eta_{\mu\nu}.
\end{equation}

The meaning of our result may be summarized saying that in the absence of gravitational sources the two metrics coincide and we have only one flat metric: Minkowski. In fact this is a simplicity criterion since we expect to recover the results of special relativity in the absence of gravitation. Take, for example, our energy-momentum conservation condition (\ref{new conservation}), if the background metric is Minkowski, when the dynamical metric is Minkowski too, we get naturally the energy conservation as given in special relativity:
\begin{equation}
\partial_\nu(T^{\mu\nu})=0,
\end{equation}
once the mass term vanishes.

If the background metric is not Minkowski the special relativity is not recovered, because the mass term would not disappear due the coupling of the two metrics.

With all these features the bases of the theory is very close to the foundations of general relativity.

\section{Cosmological consequences?}
\label{sec:6}
To illustrate the condition (\ref{new conservation}), let us consider the simple case of matter in the form of an ideal pressure-less fluid, i.e., a cloud of dust particles:
\begin{equation}
T^{\mu\nu}=\rho U^\mu U^\nu,
\end{equation}
the Robertson-Walker metric as the dynamic metric and Minkowski as the background one. Then, applying the condition (\ref{new conservation}) we have
\begin{equation}\label{rho evol}
\dot{\rho}+\left[ 3\rho+\frac{3m_g^2c^6}{16\pi G\hbar^2}(4a^2+a^4-3)\right] \frac{\dot{a}}{a}=0,
\end{equation}
and equation (\ref{cons mass}) is automatically satisfied.

Solving (\ref{rho evol}) we obtain the evolution of the energy density as a function of the scale factor:
\begin{equation}\label{dust evolution}
\rho(a)=\frac{\rho_0}{a^3}-\frac{3m_g^2c^6}{8\pi G\hbar^2}\left( \frac{a^4}{14}+\frac{2a^2}{5}-\frac{1}{2}\right),
\end{equation}
here, the first term is the evolution of the energy density as calculated in general relativity, and we have an additional term due to the mass term. This may be an interesting treatment of the mass of the graviton in cosmological scenarios, once we can interpret it like a fluid and maybe explain some observational effects that has been attributed to the cosmological constant, quintessence and other exotic fluids \cite{marcio2006}.
\\
\\

Another interesting feature emerges from our treatment. It is not possible to obtain a de Sitter solution for the vacuum.

Einstein gravity has a family of solutions given by:
\begin{equation}
G_{\mu\nu}-\Lambda g_{\mu\nu}=-\frac{8\pi G}{c^4}T_{\mu\nu}
\end{equation}
that is in accordance with the conservation laws for any small constant $\Lambda$. The vacuum solution of this equation with the Robertson-Walker metric with $k=0$, gives us the de Sitter space-time:
\begin{equation}
ds^2=dt^2-[\exp 2(\tfrac{1}{3}\Lambda)^{\frac{1}{2}}t][dr^2+r^2(d\theta^2+\sin^2\theta d\phi^2)]
\end{equation}

If we add a cosmological constant in the vacuum equations of Visser gravity:
\begin{equation}\label{eqmotion with lambda}
G_{\mu\nu}-\Lambda g_{\mu\nu}=-\frac{8\pi G}{c^4} T_{\mu\nu}^{mass},
\end{equation}
and taking the covariant divergence, from the right-hand-side we reobtain equation (\ref{vac cons}). Since the background metric is Minkowski, from (\ref{aoboa}) the dynamical metric $g_{\mu\nu}$ is Minkowski too and we obtain
\begin{equation}
G_{\mu\nu}=T_{\mu\nu}^{mass}=0,
\end{equation}
and from (\ref{eqmotion with lambda}) we have
\begin{equation}
\Lambda=0.
\end{equation}

Thus, in order to have consistency, $\Lambda$ must be rigorously zero. Since the background metric needs to be Minkowski, the cosmological vacuum solution in Visser's theory is the static flat Minkowski space-time or, e.g., some kind of cosmological parameter (like $\Lambda (t)$). For this last alternative, we would have a coupling equation like (\ref{rho evol}), which would describe the evolution of the energy density of the vacuum component.

\section{Conclusion}
\label{sec:7}

Our interpretation of the energy-momentum conservation in the Visser's massive gravity is in accordance to the equivalence principle and recover naturally the results of special relativity in the absence of gravitational sources.

The point of view considered in this paper allow us to consider Minkowski as background metric in Visser's theory in all astrophysical cases including cosmology.

This new interpretation may lead to interesting cosmological results once we can construct a cosmological model in a theory with massive gravitons with a Minkowski background. Additional contributions to the cosmological fluids will appear due to the modifications in the interaction potential, which, maybe, would be a way of treat the dark-energy problem. The analyses of the theory in the absence of gravitational sources lead us to exclude the de Sitter space-time as a vacuum solution of the massive gravity, once a constant $\Lambda$ term is rigorously zero in a flat background.

Another interesting feature is that our interpretation of the energy conservation in strong fields is independent of the form of the tensor which interact with the perfect-fluid tensor, so this can be used to other models with additional energy-momentum contribution.

\begin{acknowledgements}
MESA would like to thank the Brazilian Agency FAPESP for
support (grant 06/03158-0). ODM and JCNA would like to thank the Brazilian
agency CNPq for partial support (grants 305456/2006-7 and 303868/2004-0
respectivelly).
\end{acknowledgements}



\end{document}